\documentclass[11pt,superscriptaddress,aps,prd,preprint]{revtex4-1}
\usepackage{amsmath}
\usepackage{amssymb}
\usepackage{graphicx}
\usepackage{slashed}

\begin{document}

\title{Scattering of Fermions by Gravitons}
\author{S. C. Ulhoa }
\email{sc.ulhoa@gmail.com} \affiliation{Instituto de F\'{i}sica,
Universidade de Bras\'{i}lia, 70910-900, Bras\'{i}lia, DF,
Brazil.}

\author{A. F. Santos}
\email{alesandroferreira@fisica.ufmt.br}
\affiliation{Instituto de F\'{\i}sica, Universidade Federal de Mato Grosso,\\
78060-900, Cuiab\'{a}, Mato Grosso, Brazil.}
\affiliation{Department of Physics and Astronomy, University of Victoria,
3800 Finnerty Road Victoria, BC, Canada.}

\author{Faqir C. Khanna\footnote{Professor Emeritus - Physics Department, Theoretical Physics Institute, University of Alberta - Canada}}
\email{khannaf@uvic.ca}
\affiliation{Department of Physics and Astronomy, University of Victoria,
3800 Finnerty Road Victoria, BC, Canada.}

\date{\today}

\begin{abstract}

The interaction between gravitons and fermions is investigated in the teleparallel gravity. The scattering of fermions and gravitons in the weak field approximation is analyzed. The transition amplitudes of M$\varnothing$ller, Compton and new gravitational scattering are calculated.
\end{abstract}


%\pacs{04.20-q; 04.20.Cv; 02.20.Sv}

\maketitle
\section{Introduction}
%\noindent

In Einstein theory, the gravitational field which uses Christoffel symbols cannot couple to any fermion field. However approximations to the exact theory, like gravitoelectromagnetism (GEM) and others, have accomplished coupling to fermions and/or bosons in a cartesian quantum field theory. In such theories scattering processes between fermions and bosons have been calculated both at zero and finite temperatures \cite{Ramos2010, Khanna2016}. Teleparallel gravity is an alternative theory of gravitation in which the tetrad field is the dynamical variable. It is equivalent to general relativity, however it is constructed in terms of torsion instead of the curvature. In the framework of teleparallel gravity the coupling between fermions and gravity is accomplished by the contorsion tensor \cite{Maluf:2013gaa}. Then it is possible to study the scattering amplitude between gravitons and fermions.

Teleparallel gravity was first introduced by Einstein in an attempt to construct a unified field theory \cite{einstein}. The equivalence between two theories is achieved by identities involving the curvature scalar. There is a dynamical equivalence except for some aspects in particular a definition of the gravitational energy. The main advantage of teleparallel gravity is the fact that the gravitational energy and angular momentum are well defined. It has been applied over the years to explain the acceleration of the universe \cite{ANDP:ANDP201000168} and the general definition of entropy \cite{Maluf:2012na} as well as the role of the gravitational energy-momentum tensor \cite{Maluf:2013gaa} in general.

In particle physics the lagrangian that defines the dynamics is in cartesian coordinates. The interaction between particles is written such that dynamical processes are displayed in terms of Feynman graphs that shows propagators and interactions among particles. For instance the interaction between the Dirac field and the electromagnetic field leads to Feynman graphs for calculating scattering processes such as M$\varnothing$ller, Compton and others scattering processes \cite{opac-b1131978}.

Although the field equations of general relativity and teleparallel gravity are equivalent. For the linearized version, the graviton propagator definition is different. In general relativity the graviton propagator is defined using the scalar curvature. In teleparallel gravity the graviton propagator is obtained by writing the lagrangian density in the weak field approximation. In this article the scattering between fermions and gravitons is calculated using teleparallel gravity with weak field approximation. The process is calculated using the lagrangian density of teleparallel gravity which is invariant under global Lorentz transformation. The general relativity is invariant under local and global Lorentz transformation.

In section \ref{tel}, some details of the teleparallel gravity are recalled. In section \ref{S}, the scattering amplitude among gravitons and fermions: M$\varnothing$ller, Compton and new gravitational scattering are given. Finally in the last section some concluding remarks are presented.

{\bf Notation:} The quantities with Latin indices, $a=(0),(i)$, transform under SO(3,1) symmetry, while the ones with Greek indices, $\mu=0,i$, follow diffeomorphisms. We use $\hbar=c=G=1$.

\section{Teleparallel
Equivalent to General Relativity (TEGR)}\label{tel}

In general relativity the Christoffel symbols ${}^0\Gamma_{\mu \lambda\nu}$ and the metric tensor play central roles. It is possible to describe general relativity in terms of the tetrad field $e^{a}\,_{\nu}$ which is adapted to some observer. Such a relation is settled once $e_{(0)}\,^{\nu}$ is associated to the 4-velocity of the observer. Thus given a metric tensor there are an infinite number of tetrads, $g_{\mu\nu}=e^a\,_\mu e_{a\nu}$, each one describing a reference frame  \cite{Maluf:2013gaa}.

It is well known that Riemannian geometry is described by curvature and a vanishing torsion. On the other hand it is equivalent to a geometry, the Weitzenb\"ock space-time, described by torsion and a vanishing curvature. In order to describe such a relationship a manifold endowed with the connection $\Gamma_{\mu \lambda\nu}$, known as the Cartan connection, which is explicitly given by $\Gamma_{\mu \lambda\nu}=e_{a\mu}\partial_{\lambda} e^{a}\,_{\nu}$ is considered. It defines a torsion as

\begin{equation}
T^{a}\,_{\lambda\nu}=\partial_{\lambda} e^{a}\,_{\nu}-\partial_{\nu}
e^{a}\,_{\lambda}\,. \label{3}
\end{equation}

The Cartan connection satisfies the following identity
\begin{equation}
\Gamma_{\mu \lambda\nu}= {}^0\Gamma_{\mu \lambda\nu}+ K_{\mu
\lambda\nu}\,, \label{2}
\end{equation}
where

\begin{eqnarray}
K_{\mu\lambda\nu}&=&\frac{1}{2}(T_{\lambda\mu\nu}+T_{\nu\lambda\mu}+T_{\mu\lambda\nu})\label{3.5}
\end{eqnarray}
is the contortion tensor. Thus it is possible to write the curvature scalar $R({}^0\Gamma)$ in terms of the torsion tensor

\begin{equation}
eR({}^0\Gamma)\equiv -e({1\over 4}T^{abc}T_{abc}+{1\over
2}T^{abc}T_{bac}-T^aT_a)+2\partial_\mu(eT^\mu)\,,\label{5}
\end{equation}
where $e$ is the determinant of the tetrad field, $T_a=T^b\,_{ba}$ and $T_{abc}=e_b\,^\mu e_c\,^\nu T_{a\mu\nu}$. Then a gravitational theory equivalent to the general relativity is established using the following lagrangian density
\begin{equation}
\mathfrak{L}= -k e({1\over 4}T^{abc}T_{abc}+{1\over
2}T^{abc}T_{bac}- T^aT_a) -\mathfrak{L}_M \label{lag}\,,
\end{equation}
where $k=1/16\pi$ and $\mathfrak{L}_M$ stands for the lagrangian
density of the matter field. Here $\mathfrak{L}_M$ represents the lagrangian density of the Dirac field.

It is convenient to rewrite the lagrangian density as
\begin{equation}
\mathfrak{L}\equiv -ke\Sigma^{abc}T_{abc} -\mathfrak{L}_M\,,
\label{5.1}
\end{equation}
where

\begin{equation}
\Sigma^{abc}={1\over 4} (T^{abc}+T^{bac}-T^{cab}) +{1\over 2}(
\eta^{ac}T^b-\eta^{ab}T^c)\,. \label{6}
\end{equation}
Performing a variation with respect to the tetrad field yields the field equation

\begin{equation}
e_{a\lambda}e_{b\mu}\partial_\nu (e\Sigma^{b\lambda \nu} )- e
(\Sigma^{b\nu}\,_aT_{b\nu\mu}- {1\over 4}e_{a\mu}T_{bcd}\Sigma^{bcd}
)={1\over {4k}}eT_{a\mu}\,, \label{7}
\end{equation}
where $\delta \mathfrak{L}_M / \delta e^{a\mu}=eT_{a\mu}$. This field equation is equivalent to the Einstein equation. Then we express the field equation as
\begin{equation}
\partial_\nu(e\Sigma^{a\lambda\nu})={1\over {4k}}
e\, e^a\,_\mu( t^{\lambda \mu} + T^{\lambda \mu})\;, \label{8}
\end{equation}
where $T^{\lambda\mu}=e_a\,^{\lambda}T^{a\mu}$ and
$
t^{\lambda \mu}=k(4\Sigma^{bc\lambda}T_{bc}\,^\mu- g^{\lambda
\mu}\Sigma^{bcd}T_{bcd})\,. \label{9}
$
The quantity, $t^{\lambda \mu}$, is the gravitational energy-momentum tensor~\cite{maluf2,PhysRevLett.84.4533}. Using $t^{\lambda \mu}$, it is possible to construct a proper definition of an energy-momentum vector, by integrating over a 3-dimensional hypersurface, which is independent of coordinate transformations and dependent only on the reference frame.

\section{Graviton and Fermion scattering}\label{S}
\noindent

Here the graviton propagator for the teleparallel gravity in the weak field approximation is calculated. Then the vertex and transition amplitude for the interaction among gravitons and fermions are obtained.

\subsection{Graviton propagator}

The free lagrangian for the teleparallel gravity is given as
\begin{equation}
\mathfrak{L}_g=-ke\Sigma^{abc}T_{abc}\,,
\label{5.3}
\end{equation}
where $T_{abc}=e_b\,^\mu e_c\,^\nu T_{a\mu\nu}$ and $T_{a\mu\nu}=\partial_{\mu}e_{a\nu}-\partial_{\nu}e_{a\mu}$. Then we have
\begin{eqnarray}
\mathfrak{L}_g&=&-ke\Sigma^{abc}e_b\,^{\mu}e_c\,^{\nu}T_{a\mu\nu},\nonumber\\
&=&-2ke\Sigma^{abc}e_b\,^{\mu}e_c\,^{\nu}(\partial_{\mu}e_{a\nu}),\nonumber\\
&=&-2ke\left[\frac{1}{4}(T^{abc}+T^{bac}-T^{cab})+\frac{1}{2}(\eta^{ac}T^b-\eta^{ab}T^c)\right]e_b\,^{\mu}e_c\,^{\nu}(\partial_{\mu}e_{a\nu})\,,\label{3}
\end{eqnarray}
where eq. (\ref{6}) is used. The lagrangian i as
\begin{eqnarray}
\mathfrak{L}_g=-2ke\left[\frac{1}{4}\left({\mathfrak{L}_g}_1+{\mathfrak{L}_g}_2\right)+\frac{1}{2}{\mathfrak{L}_g}_3\right],
\end{eqnarray}
where
\begin{equation}
{\mathfrak{L}_g}_1=T^{abc}e_b\,^{\mu}e_c\,^{\nu}(\partial_{\mu}e_{a\nu})=(\partial^{\mu}e^{a\nu})(\partial_{\mu}e_{a\nu})-(\partial^{\nu}e^{a\mu})(\partial_{\mu}e_{a\nu})\,,\label{4}
\end{equation}
\begin{eqnarray}
{\mathfrak{L}_g}_2&=&(T^{bac}-T^{cab})e_b\,^{\mu}e_c\,^{\nu}(\partial_{\mu}e_{a\nu})\nonumber\\
&=&[e^a\,_{\lambda}e^c\,_{\gamma}(\partial^{\lambda}e^{b\gamma}-\partial^{\gamma}e^{b\lambda})-e^a\,_{\lambda}e^b\,_{\gamma}(\partial^{\lambda}e^{c\gamma}-\partial^{\gamma}e^{c\lambda})]e_b\,^{\mu}e_c\,^{\nu}(\partial_{\mu}e_{a\nu})\,,\nonumber\\
&=&(\partial_{\nu}e^c\,_{\gamma})(\partial^{\gamma}e_c\,^{\nu})+(\partial_{\nu}e^{c\gamma})(\partial_{\gamma}e_c\,^{\nu})+(\partial_{\lambda}e^{c\lambda})(\partial_{\mu}e_c\,^{\mu})-g_{\lambda\nu}(\partial^{\mu}e^{c\lambda})(\partial_{\mu}e_c\,^{\nu})\,,\label{5}
\end{eqnarray}
and
\begin{eqnarray}
{\mathfrak{L}_g}_3&=&(\eta^{ac}T^b-\eta^{ab}T^c)e_b\,^{\mu}e_c\,^{\nu}(\partial_{\mu}e_{a\nu})\nonumber\\
&=&e^d\,_{\lambda}(\partial^{\lambda}e_d\,^{\mu}-\partial^{\mu}e_d\,^{\lambda})
e^a\,^{\nu}(\partial_{\mu}e_{a\nu})-e^d\,_{\lambda}(\partial^{\lambda}e_d\,^{\nu}-\partial^{\nu}e_d\,^{\lambda})
e^a\,^{\mu}(\partial_{\mu}e_{a\nu})\,,\nonumber\\
&=&e^{a\nu}e_{d\lambda}(\partial^{\lambda}e^{d\mu})(\partial_{\mu}e_{a\nu})-e^{a\nu}e_{d\lambda}(\partial^{\mu}e^{d\lambda})(\partial_{\mu}e_{a\nu})+\nonumber\\
&+&e^{a\mu}e_{d\lambda}(\partial^{\nu}e^{d\lambda})(\partial_{\mu}e_{a\nu})-(\partial^{\lambda}e_{a\lambda})(\partial_{\mu}e^{a\mu})\,.\label{6.1}
\end{eqnarray}
Then the total lagrangian is
\begin{eqnarray}
\mathfrak{L}_g&=&-2ke\biggl[\frac{1}{4}\Big((\partial^{\mu}e^{a\nu})(\partial_{\mu}e_{a\nu})-(\partial^{\nu}e^{a\mu})(\partial_{\mu}e_{a\nu})+(\partial_{\nu}e^c\,_{\gamma})(\partial^{\gamma}e_c\,^{\nu})+(\partial_{\nu}e^{c\gamma})(\partial_{\gamma}e_c\,^{\nu})+\nonumber\\
&+&(\partial_{\lambda}e^{c\lambda})(\partial_{\mu}e_c\,^{\mu})-g_{\lambda\nu}(\partial^{\mu}e^{c\lambda})(\partial_{\mu}e_c\,^{\nu})\Big)+\frac{1}{2}\Big(e^{a\nu}e_{d\lambda}(\partial^{\lambda}e^{d\mu})(\partial_{\mu}e_{a\nu})-e^{a\nu}e_{d\lambda}(\partial^{\mu}e^{d\lambda})(\partial_{\mu}e_{a\nu})+\nonumber\\
&+&e^{a\mu}e_{d\lambda}(\partial^{\nu}e^{d\lambda})(\partial_{\mu}e_{a\nu})-(\partial^{\lambda}e_{a\lambda})(\partial_{\mu}e^{a\mu})\Big)\biggr]\,.\label{7}
\end{eqnarray}
Using the relations
\begin{eqnarray}
\partial_{\mu}e&=&e\,e_a\,^{\nu}\partial_{\mu}e^a\,_{\nu}\,,\\
e^{a\mu}(\partial_{\lambda}e)g^{\lambda\nu}(\partial_{\mu}e_{a\nu})&=&e^{a\mu}(\partial_{\lambda}e)(\partial_{\mu}e_{a}\,^{\lambda})-(\partial_{\lambda}e)(\partial_{\nu}g^{\lambda\nu})\,,\\
(\partial_{\nu}e^{c\gamma})(\partial_{\gamma}e_c\,^{\nu})&=&e^{c}\,_{\lambda}(\partial_{\nu}g^{\gamma\lambda})(\partial_{\gamma}e_c\,^{\nu})+(\partial_{\nu}e_{c\lambda})(\partial^{\lambda}e^{c\nu})\,,\\
g_{\lambda\nu}(\partial^{\mu}e^{c\lambda})(\partial_{\mu}e_c\,^{\nu})&=&-e_{c}\,^{\lambda}(\partial^{\mu}g_{\lambda\nu})(\partial_{\mu}e^{c\nu})+(\partial^{\mu}e_{c\nu})(\partial_{\mu}e^{c\nu})\,,\\
(\partial_{\lambda}e^{c\lambda})&=&e^{c}\,_{\nu}(\partial_{\lambda}g^{\nu\lambda})+(\partial^{\lambda}e^{c}\,_{\lambda})\,,
\end{eqnarray}
the lagrangian becomes
\begin{eqnarray}
\mathfrak{L}_g&=&-2ke\biggl[\frac{1}{4}\Big((\partial_{\nu}e^c\,_{\gamma})(\partial^{\gamma}e_c\,^{\nu})-(\partial^{\lambda}e^a\,_{\lambda})(\partial_{\mu}e_a\,^{\mu})+e^{c}\,_{\lambda}(\partial_{\nu}g^{\gamma\lambda})(\partial_{\gamma}e_c\,^{\nu})+\nonumber\\
&+&e^{c}\,_{\nu}(\partial_{\lambda}g^{\nu\lambda})(\partial_{\mu}e_c\,^{\mu})+e_{c}\,^{\lambda}(\partial^{\mu}g_{\lambda\nu})(\partial_{\mu}e^{c\nu})\Big)+\frac{1}{2e}\Big((\partial_{\lambda}e)(\partial_{\nu}g^{\lambda\nu})+\frac{1}{e}(\partial^{\mu}e)(\partial_{\mu}e)\Big)\biggr]\,.\label{9}
\end{eqnarray}

In the weak field approximation
\begin{equation}
g_{\mu\nu}=\eta_{\mu\nu}+h_{\mu\nu},\label{wa}
\end{equation}
we get
\begin{equation}
\mathfrak{L}_g\simeq e_{c\lambda}D^{ac\lambda\gamma}e_{a\gamma},
\end{equation}
where
\begin{equation}
D^{ac\lambda\gamma}=k\eta^{ac}\partial^\gamma\partial^\lambda.
\end{equation}
The graviton propagator, $\Delta_{bd\lambda\gamma}$, is obtained using the equation
$$
D^{ac\lambda\gamma}\Delta_{bd\lambda\gamma}=\frac{i}{2}\left(\delta^a_b\delta^c_d+\delta^c_b\delta^a_d\right),
$$
\begin{equation}
\Delta_{bd\lambda\gamma}=\frac{i\eta_{bd}}{kq^\lambda q^\gamma}.
\end{equation}
This propagator is represented in FIG. 1. The propagator in linearized general relativity is $D_{\mu\nu\alpha\beta}=i\left(\frac{\eta_{\mu\alpha}\eta_{\nu\beta}+\eta_{\mu\beta}\eta_{\nu\alpha}-\eta_{\mu\nu}\eta_{\alpha\beta}}{2q^2}\right)$. It is clear that a more complicated form is due to the fact that the divergence part is dropped to obtain the lagrangian density of the teleparallel gravity.
\begin{figure}[h]
\includegraphics[scale=0.7]{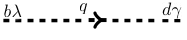}
\caption{Graviton Propagator}
\end{figure}

\subsection{Graviton-Fermion vertex}

The lagrangian for the Dirac field is given as
\begin{equation}
\mathfrak{L}_M=e\left[\frac{i}{2}\left(\bar{\Psi}\gamma^\mu D_\mu\Psi-D_\mu\bar{\Psi}\gamma^\mu\Psi\right)-m\bar{\Psi}\Psi\right],
\end{equation}
where the  covariant derivative is
\begin{equation}
D_\mu\Psi=\partial_\mu\Psi+\frac{i}{4}K_{\mu ab}\Sigma^{ab}\Psi
\end{equation}
with $K_{\mu ab}$ being the contortion tensor, $\Sigma^{ab}=\frac{i}{2}[\gamma^a,\gamma^b]$ and $m$ is the mass. The interaction lagrangian becomes
\begin{equation}
\mathfrak{L}_i=-\frac{e}{8}\left(\bar{\Psi}\gamma^\mu K_{\mu ab}\Sigma^{ab}\Psi+\bar{\Psi}K_{\mu ab}\Sigma^{ab}\gamma^\mu\Psi\right).
\end{equation}
Using
\begin{equation}
K_{\mu ab}\Sigma^{ab}=e_b\,^\nu\left(\partial_\mu e_{a\nu}-\partial_\nu e_{a\mu}\right)\Sigma^{ab}+e_b\,^\nu g_{\mu\lambda}\partial_\nu e_a\,^{\lambda}\Sigma^{ab},
\end{equation}
the lagrangian is written as
\begin{equation}
\mathfrak{L}_i=\bar{\Psi}e_{b\sigma}V^{ab\sigma\nu}e_{a\nu}\Psi,
\end{equation}
where
\begin{equation}
V^{ab\sigma\nu}=\frac{ie}{8}\Sigma^{ab}\eta^{\sigma\nu}\left(\gamma^\mu q_{2\mu}+q_{2\mu}\gamma^\mu\right),
\end{equation}
is the vertex of the interaction between two gravitons and two fermions as represented in FIG.2.
\begin{figure}[h]
\includegraphics[scale=0.5]{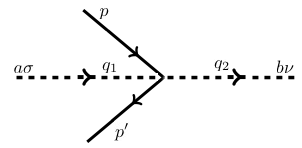}
\caption{Graviton-Fermion Vertex}
\end{figure}

In order to write the interaction vertex between two fermions and one graviton we use the approximation
\begin{equation}
e_{a\nu}=\delta_{a\nu}+\phi_{a\nu},
\end{equation}
where $\delta_{a\nu}$ is the Kronecker delta and $\phi_{a\nu}$ is a weak tetrad field. Thus
\begin{equation}
\mathfrak{L}_i=\bar{\Psi}\left(\delta_{a\nu}+\phi_{a\nu}\right)V^{ab\sigma\nu}\left(\delta_{b\sigma}+\phi_{b\sigma}\right)\Psi.
\end{equation}
Using $\Sigma_\lambda\,^\lambda=0$ since $\left[\gamma_\sigma, \gamma^\sigma\right]=0$, we get
\begin{equation}
\mathfrak{L}_i=\bar{\Psi}{\cal V}^{b\sigma}\phi_{b\sigma}\Psi,
\end{equation}
where ${\cal V}^{b\sigma}$ is the vertex given as
\begin{equation}
{\cal V}^{b\sigma}=\frac{ie}{4}\Sigma^{b\sigma}\left(\gamma^\mu q_\mu+q_\mu\gamma^\mu\right).
\end{equation}
This vertex is represented in FIG.3.
\begin{figure}[h]
\includegraphics[scale=0.6]{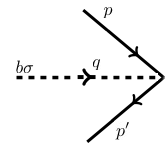}
\caption{Graviton-Fermion Vertex}
\end{figure}

To calculate processes involving fermions and gravitons the fermion propagator is necessary. It is given as
\begin{eqnarray}
S_0(p)=i\frac{\slashed p+m}{p^2-m^2},
\end{eqnarray}
and represented in FIG. 4.
\begin{figure}[h]
\includegraphics[scale=0.6]{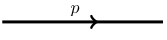}
\caption{Fermion Propagator}
\end{figure}

\subsection{Transition amplitude}

The transition amplitude, $-i{\cal M}$, is calculated for processes involving gravitons and fermions.

\subsubsection{M$\varnothing$ller scattering}

The gravitational M$\varnothing$ller scattering is given in FIG. 5.
\begin{figure}[h]
\includegraphics[scale=0.5]{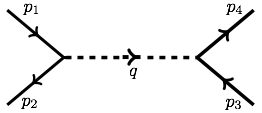}
\caption{ M$\varnothing$ller scattering}
\end{figure}

The transition amplitude for this process is
\begin{equation}
-i{\cal M}=\bar{u}(p_2){\cal V}^{b\sigma}u(p_1)\Delta_{bd\sigma\gamma}\bar{u}(p_4){\cal V}^{d\gamma}u(p_3),
\end{equation}
where ${\cal V}^{b\sigma}$ is the vertex and $\Delta_{bd\sigma\gamma}$ is the graviton propagator. Therefore the transition amplitude for this scattering is
\begin{equation}
{\cal M}=\frac{e^2}{16k}\bar{u}(p_2)u(p_1)\left(\gamma^\mu q_{\mu}+q_{\mu}\gamma^\mu\right)\frac{\Sigma^{\sigma b}\Sigma^{\gamma}\,_b}{q^\sigma q^\gamma}\left(\gamma^\nu q_{\nu}+q_{\nu}\gamma^\nu\right)\bar{u}(p_4)u(p_3).
\end{equation}

\subsubsection{Compton scattering}

The gravitational Compton scattering is given in FIG. 6.
\begin{figure}[h]
\includegraphics[scale=0.5]{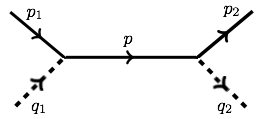}
\caption{Compton scattering}
\end{figure}

The transition amplitude for this process is
\begin{equation}
-i{\cal M}=\bar{u}(p_2){\cal V}^{b\sigma}S_0(p)\eta_{bc}\,\eta_{\sigma\nu}{\cal V}^{c\mu}u(p_1),
\end{equation}
then
\begin{equation}
{\cal M}=\frac{e^2}{16}\bar{u}(p_2)\Sigma^{b\sigma}\left(\gamma^\mu q_{1\mu}+q_{1\mu}\gamma^\mu\right)\left(\frac{\slashed p+m}{p^2-m^2}\right)\eta_{bc}\eta_{\sigma\mu}\Sigma^{c\mu}\left(\gamma^\alpha q_{2\alpha}+q_{2\alpha}\gamma^\alpha\right)u(p_1).
\end{equation}

\subsubsection{New gravitational scattering}

In the framework of teleparallelism, another scattering process arises, as given in FIG. 7.
\begin{figure}[h]
\includegraphics[scale=0.5]{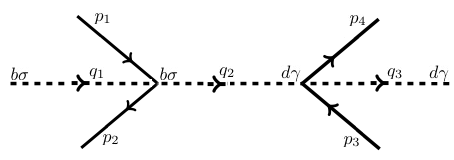}
\caption{New gravitational scattering}
\end{figure}
For this process we have
\begin{equation}
-i{\cal M}=\bar{u}(p_2){\cal V}^{b\sigma}u(p_1)\Delta_{bd\sigma\gamma}\bar{u}(p_4){\cal V}^{d\gamma}u(p_3),
\end{equation}
which yields
\begin{equation}
{\cal M}=\frac{e^2}{16k}\bar{u}(p_2)\left(\gamma^\mu q_{1\mu}+q_{1\mu}\gamma^\mu\right)u(p_1)\frac{\Sigma^{\sigma b}\Sigma^{\gamma}\,_b}{q_2^\sigma q_2^\gamma}\bar{u}(p_4)\left(\gamma^\nu q_{3\nu}+q_{3\nu}\gamma^\nu\right)u(p_3).
\end{equation}
This expression is used to calculate the scattering cross section between fermion and graviton.

The scattering matrix is $S=1-iT$, where $T\rightarrow{\cal M}$. The cross section for M$\varnothing$ller, Compton and new gravitational scattering is easily calculated.  It's interesting to note that in GEM theory the vertex shows a graviton and two fermions. Here there is a pre-existing graviton which modifies the momenta of the fermions whose encounter yields another graviton.

\section{Conclusion}

The scattering amplitude between fermions and gravitons is calculated using weak field approximation. One of the gravitons in the interaction vertex takes into account the torsion of the space-time. It is important to point out that the propagator obtained in teleparallel gravity is different from the one in general relativiy. This is due to the difference between symmetries of two theories. The scalar curvature is invariant under local and global Lorentz transformation. On the other hand teleparallel gravity is invariant under only global transformation. The Christoffel symbols are linked to contorsion, then the interaction between gravitons and photons assumes the same form as in general relativity. Using this formalism it is not possible to establish an interaction between gravitons and fermions with photons.

\section*{Acknowledgments}

This work by A. F. S. is supported by CNPq projects 476166/2013-6 and 201273/2015-2. S. C. U. thanks the Funda\c{c}\~ao de Apoio $\grave{a}$ Pesquisa do Distrito Federal - FAPDF for financial support. We thank Physics Department, University of Victoria for access to facilities.

%\bibliography{ref}
%\bibliographystyle{apsrev4-1}

%

\end{document}